\documentclass[twocolumn,showpacs,preprintnumbers,amsmath,amssymb,nofootinbib,floatfix]{revtex4}

\usepackage[dvips]{graphicx}
\usepackage{dcolumn}
\usepackage{bm}
\usepackage{braket}
\usepackage{comment}

\def\Journal#1#2#3#4{{#1} {\bf #2}, #3 (#4)}

\def\AandA{Astron. Astrophys.}

\def\CPC{Chin. Phys. C}

\def\IJMPD{Int. J. Mod. Phys. D}

\def\JCAP{J. Cosmol. Astropart. Phys.}
\def\JHEP{J. High Energy Phys.}

\def\MNRAS{Mon. Not. R. Astron. Soc.}

\def\PDU{Phys. Dark. Univ.}
\def\PLB{{Phys. Lett.} B}

\def\PRL{Phys. Rev. Lett.}
\def\PRD{Phys. Rev. D}

\begin{document}


\title{Generalized hybrid natural inflation}

\author{Teruyuki Kitabayashi}
\email{teruyuki@tokai-u.jp}
\affiliation{%
Department of Physics, Tokai University, 4-1-1 Kitakaname, Hiratsuka, Kanagawa 259-1292, Japan
}



\begin{abstract}
Although the hybrid natural inflation is a successful inflation model, the symmetry breaking scale in the model should be large for a negative value of the running of the scalar spectral index. This study proposed a generalized hybrid natural inflation model that can realize a low-scale inflation with a negative value of the running of the scalar spectral index. 
\end{abstract}

\maketitle



\section{Introduction\label{section:introduction}}
The inflationary paradigm \cite{Guth1981PRD,Sato1981MNRAS,Linde1982PLB,Albrecht1982PRL} constitutes a major portion of standard cosmology, and several inflation models have been developed, which are consistent with observations \cite{Martin2014PDU}. 

In particular, the natural inflation (NI) model \cite{Freese1990PRL,Adams1993PRD,Freese2008IJMPD,Freese2015JCAP} is attractive because the origin of the inflaton potential is well-motivated. The inflaton potential in the NI model is expressed as
\begin{eqnarray}
V(\phi)=\Lambda^4 \left[1+\cos\left(\frac{\phi}{f}\right) \right],
\label{Eq:V_NI}
\end{eqnarray}
where $\phi$ denotes the inflaton (a pseudo-Goldstone boson), $\Lambda$ indicates the inflation scale, and $f$ represents the symmetry breaking scale associated with a spontaneously symmetry breaking of a global symmetry of the inflaton. Unfavorably, the NI model is inconsistent with the recent measurement of the scalar spectral index and tensor-to-scalar ratio \cite{Planck2020AA,BICEP22018PRL}. Moreover, apart from this inconsistency, a large symmetry breaking scale $f \gtrsim 5.4 M$ is required in the NI model, where $M=1/\sqrt{8\pi G}\simeq 2.435\times 10^{18}$ GeV is the reduced Planck mass ($G$ denotes the gravitational constant). The large symmetry breaking scale is theoretically undesired because it may cause large gravitational corrections to the potential. Therefore, new inflation models have been proposed based on the NI model. 

The hybrid natural inflation (HNI) model \cite{Ross2010PLB,Ross2010PLB2,Hebecker2013PRD,Gonzalez2014PLB,Carone2014PRD,Vazquez2015JCAP,Ross2016JHEP,German2017JCAP,Ali2017PRD,Gong2021PRD} is one of the extended versions of the NI model, and its inflaton potential is expressed as
\begin{eqnarray}
V(\phi)=\Lambda^4 \left[1+a \cos\left(\frac{\phi}{f}\right) \right],
\label{Eq:V_HNI}
\end{eqnarray}
where $a$ represents the model parameter $(0<a<1)$. The HNI is classified as a hybrid model in which a second field is responsible for terminating the inflation. Utilizing this hybrid feature, a low-scale inflation at $f \simeq M$ can be realized if the value of the running of the scalar spectral index is not restricted. However, to ensure a negative value of the running of the scalar spectral index, the symmetry breaking scale should be $f \ge 3.83 M$ in the HNI model. Although a positive value of the running of the scalar spectral index is permissible in the observations, its negative value is favored. 

In this paper, a new extended NI model, generalized hybrid natural inflation (GHNI) model is proposed with an inflaton potential of
\begin{eqnarray}
V(\phi)=\Lambda^4 \left[1+a\cos\left(\frac{\phi}{f}\right) \right]^n,
\label{Eq:V_GHNI}
\end{eqnarray}
where $n$ denotes the model parameter. Although the theoretical origin of $n$ is a critical consideration, this study disregarded it as a norm in generalized model building \cite{Munoz2015PRD,Zhang2020CPC}. Objectively, the GHNI model aims to realize a low-scale inflation at $f \simeq 0.05 M$, which is consistent with the observed scalar spectral index and tensor-to-scalar ratio as well as the negative value of the running of the scalar spectral index. 

The rest of this paper is organized as follows. In Section \ref{section:slow-roll}, we present a review of slow-roll parameters and the observables related to the inflation. In Section \ref{section:GHNI}, the new inflation model, GHNI, is proposed. Herein, we establish the requirement of a large symmetry breaking scale in the HNI model to achieve a negative value of the running of the scalar spectral index. Interestingly, this negative running value of the scalar spectral index can be achieved even with a small symmetry breaking scale using the GHNI model. Finally, the present findings are summarized in Section \ref{section:summary}.

\section{Slow-roll parameters and observables \label{section:slow-roll}}
Herein, we perform the slow-roll approximation of the inflation potential. The consistency of inflation models can be estimated by slow-roll parameters, which are relevant for this study and are defined as \cite{Liddle2000}:
\begin{eqnarray}
\epsilon&=&\frac{M^2}{2} \left(\frac{V'(\phi)}{V(\phi)}\right)^2,
\label{Eq:epsilon} \\
\eta&=&M^2 \frac{V''(\phi)}{V(\phi)},\\
\label{Eq:eta}
\xi_2&=&M^4 \frac{V'(\phi)V'''(\phi)}{V^2(\phi)},\\
\label{Eq:xi2}
\xi_3&=&M^6 \frac{V'^2(\phi)V''''(\phi)}{V^3(\phi)},
\label{Eq:xi3}
\end{eqnarray}
where the prime sign denotes the derivative with respect to $\phi$.

The observables are related to the slow-roll parameters as follows:
\begin{eqnarray}
n_{\rm s}&=&1-6\epsilon+2\eta, \label{Eq:ns} \\
n_{\rm sk}&=&\frac{dn_{\rm s}}{d\ln k}=16\epsilon \eta - 24 \epsilon^2 -2\xi_2,\label{Eq:nsk} \\
n_{\rm skk}&=&\frac{d^2n_{\rm s}}{d\ln k^2} \label{Eq:nskk}\\
&=&-192\epsilon^3+192 \epsilon^2 \eta - 32 \epsilon \eta^2 -24 \epsilon\xi_2 + 2\eta \xi_2 +2 \xi_3,\nonumber \\
A_{\rm s} &=&\frac{2V}{3 \pi^2 M^4 r}, \label{Eq:As} 
\end{eqnarray}
and
\begin{eqnarray}
n_{\rm t}&=&-2\epsilon,\label{Eq:nt}\\
n_{\rm tk}&=&\frac{dn_{\rm t}}{d\ln k}=4\epsilon (\eta-2\epsilon),\label{Eq:ntk}\\
r&=&16\epsilon,\label{Eq:r}
\end{eqnarray}
where $n_{\rm s}$ denotes the scalar spectral index, $n_{\rm sk}$ indicates the running of the scalar spectral index ($k$ denotes wave number), $n_{\rm skk}$ represents the running of running of the scalar spectral index, $A_{\rm s}$ denotes the scalar power spectrum amplitude, $n_{\rm t}$ indicates the tensor spectral index, $n_{\rm tk}$ represents the running of the tensor spectral index, and $r$ indicates the tensor-to-scalar ratio, respectively. Accordingly, the parameter was defined as
\begin{eqnarray}
\delta_{\rm ns} = 1-n_{\rm s}.
\label{Eq:delta}
\end{eqnarray}

In principle, the correct inflation models should be consistent with the observation. The scalar spectral index, running scalar spectral index, scalar power spectrum amplitude, and tensor-to-scalar ratio were constrained from the observation as $n_{\rm s} = 0.9658 \pm 0.0040$, $n_{\rm sk}=-0.0066 \pm 0.0070$, $A_{\rm s} \simeq e^{3.08}\times 10^{-10} \simeq 2.2\times 10^{-9}$, and  $r < 0.068$ with Planck TT, TE, EE+lowE+lenging+BK15+BAO \cite{Planck2020AA}. In addition, the tensor-to-scalar ratio may be constrained to $r < 0.035$  \cite{BICEP22018PRL}.

For the proposed inflation model, the constraints $n_{\rm s}$, $n_{\rm sk}$, and $r$ are required.
\begin{eqnarray}
&& n_{\rm s} = 0.966 \quad (\delta_{\rm ns}=0.034), \nonumber \\
&& -0.0136 \le  n_{\rm sk}  < 0.0004, \nonumber \\
&& r \le 0.06.
\end{eqnarray}
Although a positive $n_{\rm sk}$ is still permissible in the Planck data, a negative $n_{\rm sk}$ is favored. Thus, we estimate the allowed model parameters for $n_{\rm sk}  \le 0.0004$ as well as $n_{\rm sk}  \le 0$.

\section{Generalized hybrid natural inflation \label{section:GHNI}}

\subsection{Slow-roll parameters}
The slow-roll parameters for the GHNI model defined by the potential in Eq. (\ref{Eq:V_GHNI}) are stated as follows:

\begin{eqnarray}
\epsilon&=&\frac{1}{2}\left(\frac{M}{f}\right)^2  \frac{a^2 n^2\left(1-c_\phi^2\right)}{\left(1+ac_\phi\right)^2}, \label{Eq:GHNI_epsilon}\\
\eta&=&-\left(\frac{M}{f}\right)^2  \frac{an\left[a +c_\phi+an(c_\phi^2-1)\right]}{\left(1+ac_\phi\right)^2},\label{Eq:GHNI_eta}
\end{eqnarray}
and
\begin{eqnarray}
\xi_2&=&-2\left(\frac{M}{f}\right)^2 \epsilon \label{Eq:GHNI_xi2} \\
&&\times \frac{ 1-2a^2 +a^2n(3-n) + a(3n-1)c_\phi+a^2n^2c_\phi^2 }{\left(1+ac_\phi \right)^2},\nonumber \\
\xi_3&=&-2\left(\frac{M}{f}\right)^2 \epsilon \eta \frac{c_\phi + a (a^2 x_1+ a c_\phi x_2  +x_3 ) }{(1+a c_\phi)^2\left[a+c_\phi + an(c_\phi^2-1)\right]}, \label{Eq:GHNI_xi3} \nonumber \\ 
\end{eqnarray}
where 
\begin{eqnarray}
c_\phi = \cos \left( \frac{\phi}{f} \right),
\end{eqnarray}
and
\begin{eqnarray}
x_1&=&  (n-3)(n-2)(n-1)+c_\phi^4n^3 \nonumber \\
&& -2c_\phi^2(n-1)(n(n-2)+2), \nonumber \\ 
x_2&=& 6  (c_\phi^2-1) n^2 +2(5-2c_\phi^2)n +c_\phi^2-4, \nonumber \\ 
x_3&=&4-4n+c_\phi^2(7n-4). 
\end{eqnarray}
From Eqs. (\ref{Eq:r}) and (\ref{Eq:GHNI_epsilon}), $c_\phi$ can be derived as
\begin{eqnarray}
c_\phi = \frac{-f^2r \pm 8M^2an^2\sqrt{1-\frac{(1-a^2)f^2 r}{8M^2a^2n^2}}}{a\left(f^2r+8M^2n^2\right)},
\label{Eq:GHNI_c_phi_r}
\end{eqnarray}
and from Eqs.  (\ref{Eq:GHNI_eta}) and (\ref{Eq:GHNI_c_phi_r}), we obtained 
\begin{eqnarray}
\eta &=& 
\begin{cases}
-\frac{1}{2}\left(\frac{M}{f}\right)^2 n + \frac{1}{16}\left( 2-\frac{1}{n}\right)r & (a=1) \\
\left(\frac{M}{f}\right)^2 \frac{a^2 n}{1-a^2} A & (0<a<1)
\end{cases},
\label{Eq:GHNI_eta_r}
\end{eqnarray}
where
\begin{eqnarray}
A=1 +\frac{(1-a^2)f^2(n-1)r}{8M^2a^2n^2} \pm \frac{1}{a}\sqrt{1-\frac{(1-a^2)f^2r}{8M^2a^2n^2} }.
\label{Eq:GHNI_eta_r_A} \nonumber \\
\end{eqnarray}

The symmetry breaking scale is expressed as
\begin{eqnarray}
f = \begin{cases}
\frac{2Mn}{\sqrt{4n\delta_{\rm ns}-\frac{(n+1)r}{2}}} & (a=1) \\
\frac{4Man}{\sqrt{r-a^2(n+2)r+8a^2n\delta_{\rm ns} + \sqrt{B}}} & (0<a<1)
\end{cases},
\label{Eq:GHNI_f}
\end{eqnarray}
from Eqs. (\ref{Eq:ns}), (\ref{Eq:r}), (\ref{Eq:delta}), and (\ref{Eq:GHNI_eta_r}), where
\begin{eqnarray}
B =  (1+a^2n(n+2))r^2-16a^2n(n+1)r\delta_{\rm ns}  +64 a^2n^2\delta_{\rm ns}^2   \label{Eq:GHNI_f_X}. \nonumber \\
\end{eqnarray}

Note that, for $n=1$, the equations of the GHNI model reduce to those of the HNI model. Thus, the HNI model is a specific case of the GHNI model (both are hybrid inflation models). 

We comment about the generalized natural inflation (GNI) model \cite{Munoz2015PRD,Zhang2020CPC} with the inflaton potential of
\begin{eqnarray}
V(\phi)=2^{1-n}\Lambda^4 \left[1+\cos\left(\frac{\phi}{f}\right) \right]^n. 
\label{Eq:V_GNI}
\end{eqnarray}
The equations of the GHNI model obtained with $a=1$ are identical to those of the GNI model; however, similar to the NI model, the GNI model is regarded as a single-field inflation model in which the inflation and the end of inflation are controlled by the same inflaton. In contrast, the GHNI model is a hybrid model, a second field is responsible for the end of inflation. Thus, the GNI (single-field model) is not a specific case of the GHNI (hybrid inflation model). 

Moreover, the equations of the GHNI model obtained with $n=1$ and $a=1$ are identical to those of the NI model in case a positive sign `$+$' is selected for the second term in the numerator in Eq. (\ref{Eq:GHNI_c_phi_r}) and negative sign `$-$' for the third term in the Eq.(\ref{Eq:GHNI_eta_r_A}). However, the NI (single-field model) is not a specific case of GHNI (hybrid inflation model).

\subsection{Lower limit of the symmetry breaking scale  \label{section:minimal_fl}}
%
\begin{figure}[t]
\begin{center}
\includegraphics[scale=1.0,pagebox=cropbox,clip]{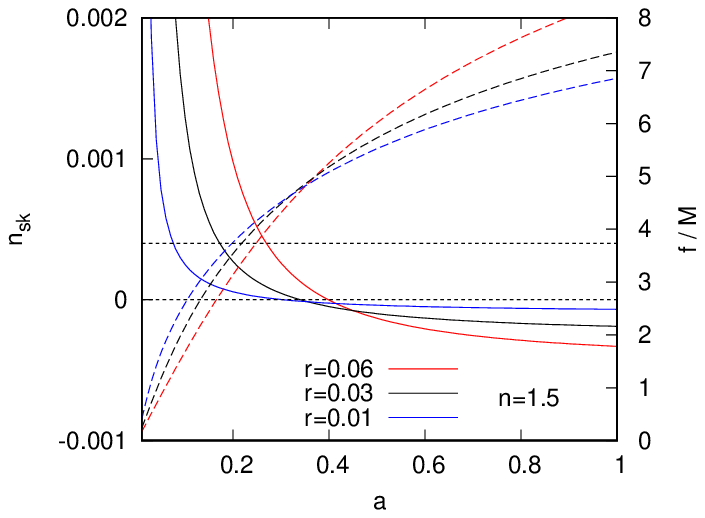}
\includegraphics[scale=1.0,pagebox=cropbox,clip]{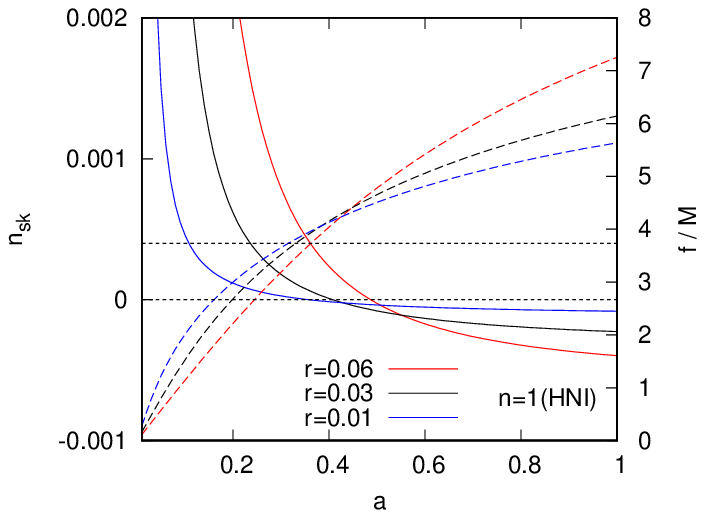}
\includegraphics[scale=1.0,pagebox=cropbox,clip]{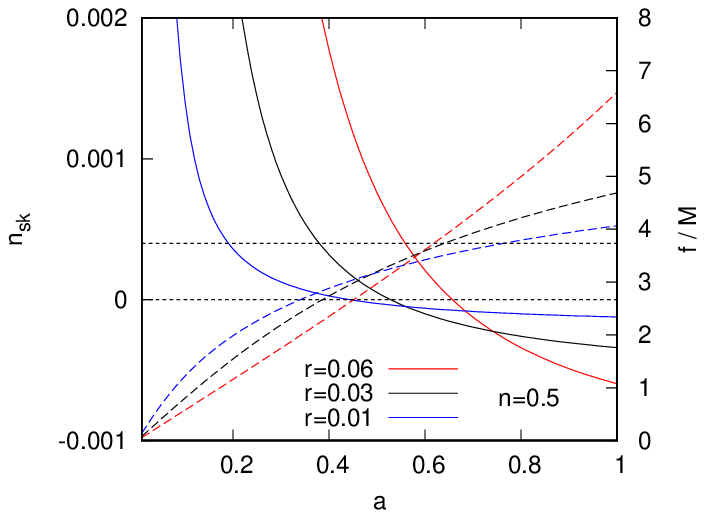}
\caption{Running of scalar spectral index $n_{\rm sk}$ (solid curves) and symmetry breaking scale $f$ (dashed curves) as a function of parameter $a$ for $r=0.01, 0.03, 0.06$. Upper and lower horizontal dotted lines in each panels correspond to $n_{\rm sk}=0.0004$ and $n_{\rm sk}=0$, respectively; middle panel ($n=1$) represents to HNI model; top and bottom panels exhibit the example of GHNI model for $n=1.5$ and $n=0.5$, respectively. }
\label{Fig:nsk_a_f} 
\end{center}
\end{figure}
\begin{figure*}[t]
\begin{center}
\includegraphics[scale=1.0,pagebox=cropbox,clip]{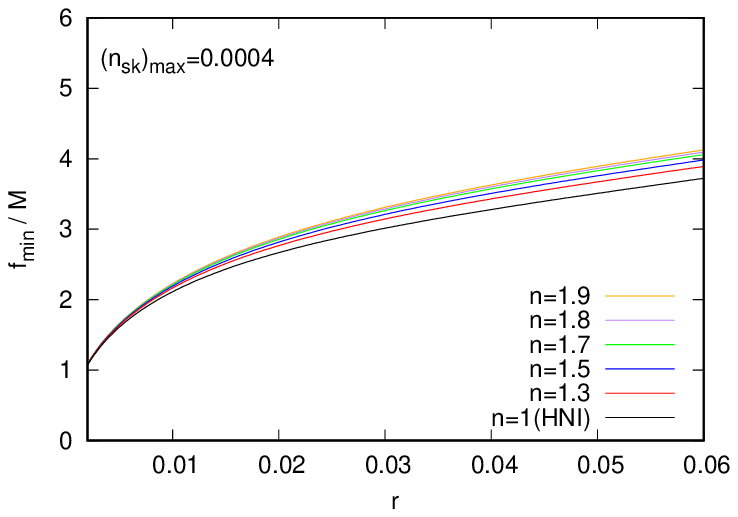}
\includegraphics[scale=1.0,pagebox=cropbox,clip]{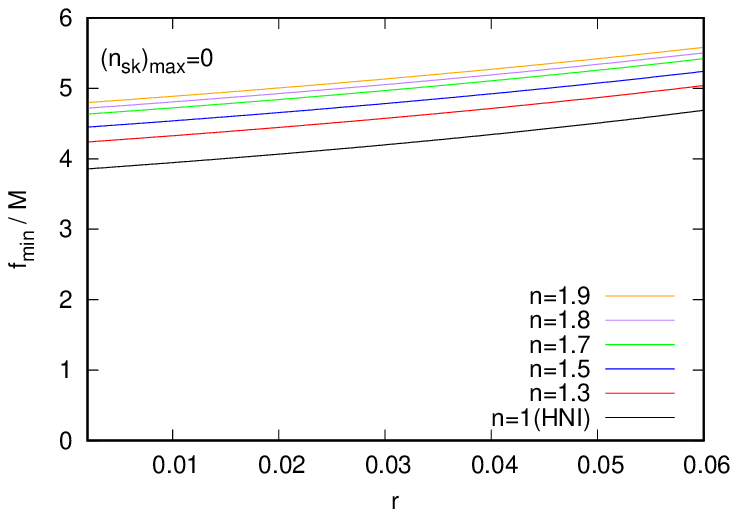}\\
\includegraphics[scale=1.0,pagebox=cropbox,clip]{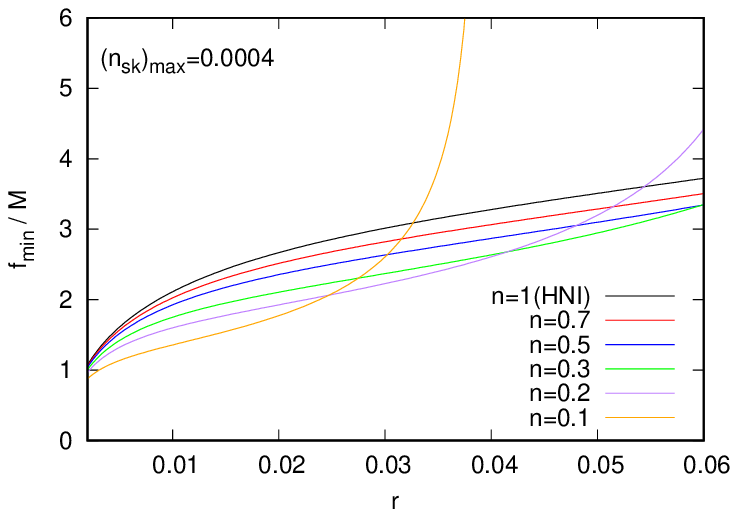}
\includegraphics[scale=1.0,pagebox=cropbox,clip]{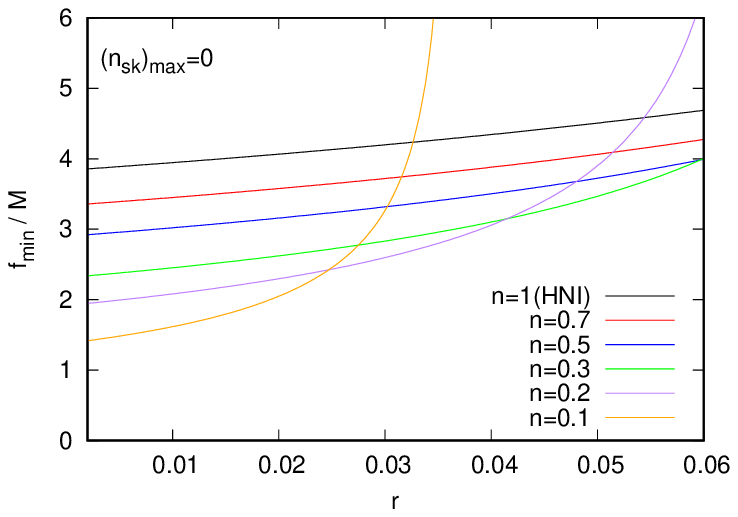}\\
\includegraphics[scale=1.0,pagebox=cropbox,clip]{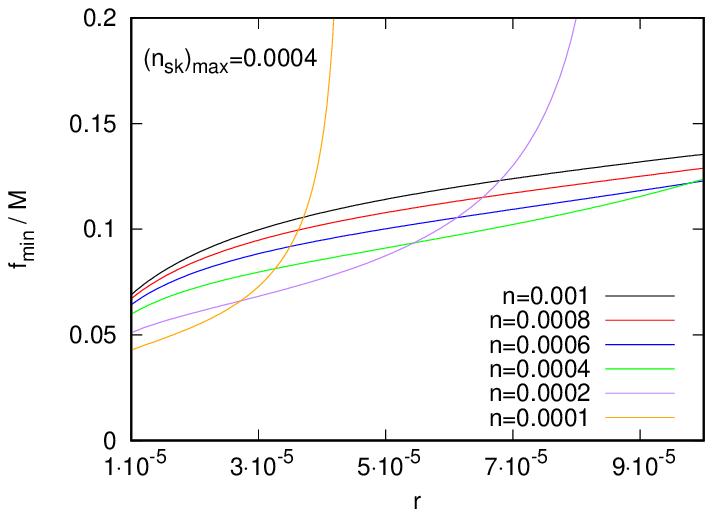}
\includegraphics[scale=1.0,pagebox=cropbox,clip]{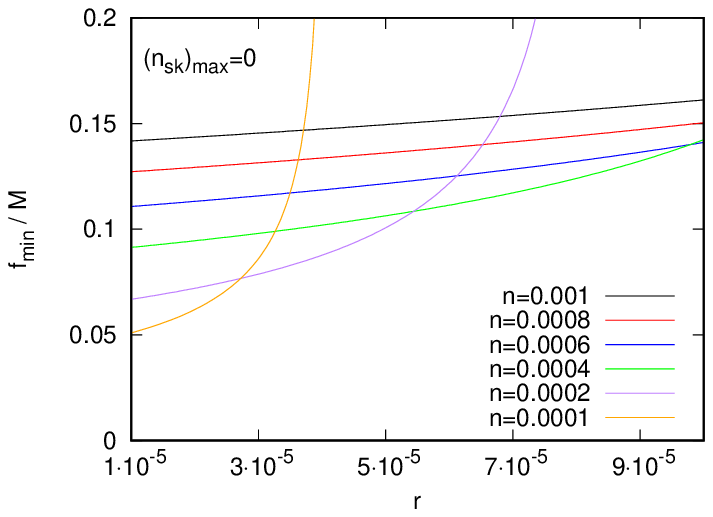}
\caption{Lower limit of symmetry breaking scale $f$ as a function of tensor-to-scalar ratio $r$ for $n=1,1.3, \cdots, 1.9$ (top panel), $n=0.1,1.2, \cdots, 1$ (middle panel), and for extremely small $n$ and $r$ (bottom panel). Three panels on left (right) display the minimum value of the symmetry breaking scale, $f_{\rm min}$, for $n_{\rm sk} \le 0.0004$ ($n_{\rm sk} \le 0$).}
\label{Fig:fmin_r} 
\end{center}
\end{figure*}

We show that the lower limit of the symmetry breaking scale in the HNI model should be $f \simeq 3.83 M$ for a negative $n_{\rm sk}$. In contrast, smaller scale $f \simeq 0.05 M$ is enabled with negative $n_{\rm sk}$ in GHNI model.

The running of the scalar spectral index $n_{\rm sk}$ (solid curves) and the symmetry breaking scale $f$ (dashed curves) is displayed in Figure \ref{Fig:nsk_a_f} as a function of parameter $a$ for $r=0.01, 0.03, 0.06$. In each panel, the upper horizontal dotted line denotes the observed upper limit $n_{\rm sk}=0.0004$, the lower horizontal dotted line exhibits $n_{\rm sk}=0$, and the middle panel ($n=1$) corresponds to the HNI model. The top and bottom panels display the example of GHNI model for $n=1.5$ and $n=0.5$, respectively.

The lower limit of the symmetry breaking scale can be estimated from Fig. \ref{Fig:nsk_a_f}. For instance, the red curve ($r=0.06$) in the top panel ($n=1.5$) intersected the dotted $n_{\rm sk}=0.0004$ line at $a = 0.27$. Thus, the symmetry breaking scale should be $f \ge 3.99 M$ for $n_{\rm sk}<0.0004$. Similarly, the red curve crossed the dotted $n_{\rm sk}=0$ line at $a = 0.40$ and the symmetry breaking scale should be $f \ge 5.25 M$, if $n_{\rm sk}<0$ is required rather than $n_{\rm sk}<0.0004$. 

The lower limit of the symmetry breaking scale $f$ is exhibited in Fig. \ref{Fig:fmin_r} as a function of the tensor-to-scalar ratio $r$ for $n=1,1.3, \cdots, 1.9$ (top panel), $n=0.1,1.2, \cdots, 1$ (middle panel), and for extremely small $n$ and $r$ (bottom panel). The three panels on the left (right) display the minimum values of the symmetry breaking scale, $f_{\rm min}$, for $n_{\rm sk} \le 0.0004$ ($n_{\rm sk} \le 0$). Notably, the symmetry breaking scale $f$ decreased with the parameter $n$ and the tensor-to-scalar ratio $r$. 

Although a small breaking scale $f\simeq M$ was enabled in the HNI model ($n=1$) for $n_{\rm sk} \le 0.0004$ (left panels in Fig. \ref{Fig:fmin_r}), we require the condition of $n_{\rm sk} \le 0$ (right panels), and thus, the lower limit of the symmetry breaking scale in the HNI model should be $f \simeq 3.83 M$. In contrast, the small symmetry breaking scale, such as $f \simeq 1.38 M$ for $n=0.1$ and $f \simeq 0.051 M$ for $n=0.0001$, was enabled with negative $n_{\rm sk}$ in the GHNI model. If the positive $n_{\rm sk}$ is excluded in future observations, the validity of the GHNI model would increase. 

Herein, we discuss the over-production of the primordial black holes in the early universe. As reported earlier, a large positive value of the running of the scalar spectrum index $n_{\rm sk}$ may possibly cause overproduction of primordial black holes at the end of inflation \cite{Kohri2008JCAP,Josan2009PRD}. The scalar power spectrum at the first order in the slow-roll parameters is expressed as 
\begin{eqnarray}
\mathcal{P}_{\rm s} &=& \frac{1}{24\pi^2M^4}\frac{V}{\epsilon} \nonumber \\
&=& A_{\rm s}\left(\frac{k}{k_{\rm H}}\right)^{(n_{\rm s}-1)+\frac{1}{2}n_{\rm sk}\ln \left(k/k_{\rm H}\right) + \cdots },
\end{eqnarray}
in terms of wave number $k$, where $k_{\rm H}$ denotes the wave number at the horizon crossing. The Taylor expansion of the power spectrum around its value at the horizon crossing is constrained by
\begin{eqnarray}
\ln \left[\frac{\mathcal{P}_{\rm s} (0)}{\mathcal{P}_{\rm s} (N_{\rm H})} \right] = (n_{\rm s} - 1) N_{\rm H} + \frac{1}{2} n_{\rm sk}N_{\rm H}^2 \le 14,
\end{eqnarray}
where $N_{\rm H} \simeq 50 - 60$ denotes the e-folding number. The undesired amplitude for perturbations at which the primordial black holes could be overproduced was $\mathcal{P}_{\rm s}(0) \simeq 10^{-3}$. This amplitude $\mathcal{P}_{\rm s}(0)$ evolved from the initial value $\mathcal{P}_{\rm s}(N_{\rm H}) \simeq 10^{-9}$, thereby yielding the upper bound $n_{\rm sk} < 10^{-2}$. 

Accordingly, we require the condition of $n_{\rm sk} \le 0.0004$ (and $n_{\rm sk} \le 0$), and this study resolved the issue related to the over-production of the primordial black holes.  

\subsection{$n_{\rm skk}$, $n_{\rm tk}$ and $\Lambda$  \label{section:nskk_ntk}}
\begin{figure}[t]
\begin{center}
\includegraphics[scale=1.0,pagebox=cropbox,clip]{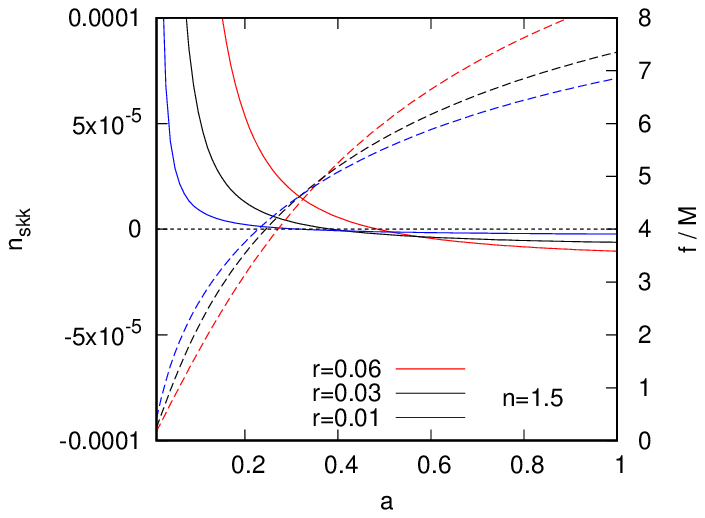}
\includegraphics[scale=1.0,pagebox=cropbox,clip]{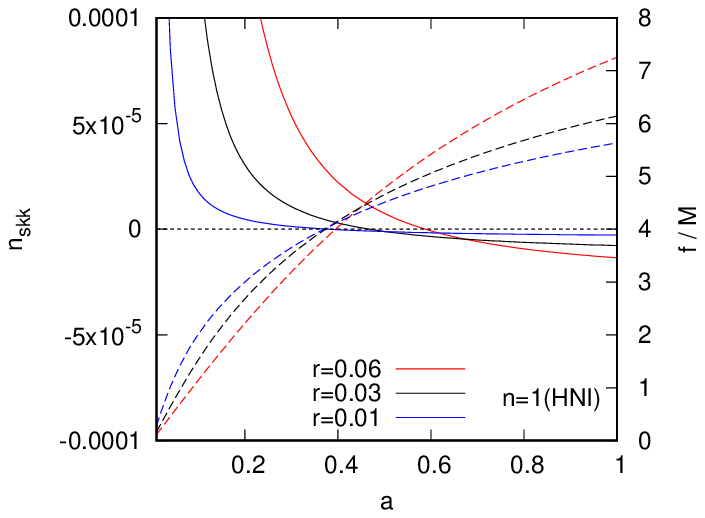}
\includegraphics[scale=1.0,pagebox=cropbox,clip]{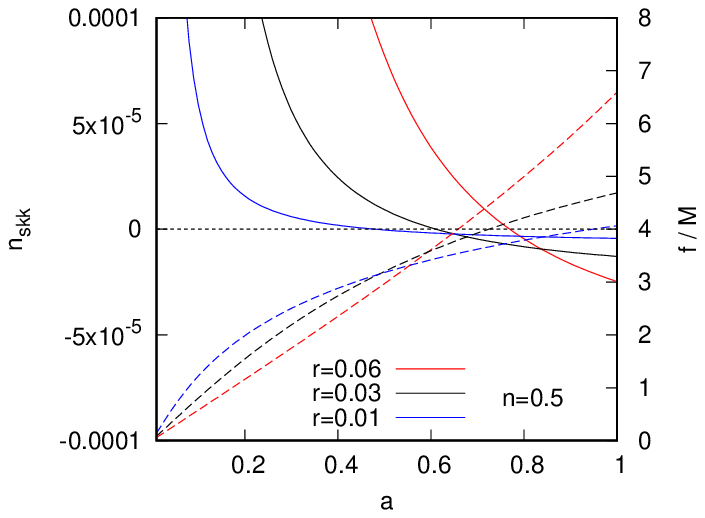}
\caption{Running of running of scalar spectral index $n_{\rm skk}$ (solid curves) and symmetry breaking scale $f$ (dashed curves) as a function of parameter $a$ for $r=0.01, 0.03, 0.06$. For reference, horizontal dotted lines denote $n_{\rm skk}=0$; middle panel ($n=1$) corresponds the HNI model; top and  bottom panels display examples of the GHNI model for $n=1.5$ and $n=0.5$, respectively.}
\label{Fig:nskk_a_f} 
\end{center}
\end{figure}

Although the running of running of the scalar spectral index $n_{\rm skk}$, the running of the tensor spectral index $n_{\rm tk}$, and the energy scale of inflation $\Lambda$ are not useful for constraining the parameters in the GHNI model, these quantities were estimated for the completeness of this study. 

The running of running of the scalar spectral index $n_{\rm skk}$ (solid curves) and the symmetry breaking scale $f$ (dashed curves) are depicted in Fig. \ref{Fig:nskk_a_f} as a function of parameter $a$ for $r=0.01, 0.03, 0.06$, wherein the horizontal dotted lines indicate $n_{\rm skk}=0$ as a reference. The middle panel ($n=1$) corresponds to the HNI model; the top and bottom panels display examples of GHNI model for $n=1.5$ and $n=0.5$, respectively.

The running of the tensor spectral index was estimated as $n_{\rm tk}=r(r-8\delta_{\rm ns})/64$. For example, we obtained $n_{\rm tk} = -1.99 \times 10^{-4}$ for $r=0.06$ and $\delta_{\rm ns}=0.0034$. 

The energy scale of inflation $\Lambda$ can be investigated using the relation between the scalar power spectrum amplitude $A_{\rm s}$ and the inflation potential in Eq. (\ref{Eq:As}), with the GHNI potential in Eq.(\ref{Eq:V_GHNI}) expressed as 
\begin{eqnarray}
\Lambda = 3.89 \times 10^{16} \left[ 2^{n-2} r (1+ac_\phi)^{-n} \right]^{1/4} \quad {\rm GeV}.
\label{Eq:GHNI_Lambda}
\end{eqnarray}
For instance, we derived
\begin{eqnarray}
\frac{\Lambda}{ 10^{16} \ {\rm GeV}} = 
\begin{cases}
1.55 - 1.71  & (n=1) \\
1.45  - 1.53  & (n=0.5) \\
1.38 - 1.39  & (n=0.1) \\
\end{cases},
\end{eqnarray}
for $a = 0.2$, $r=0.06$, and $-1 \le c_\phi \le 1$.

\section{Summary\label{section:summary}}
Although the NI model is attractive because the origin of the inflaton potential is well-motivated, this model is inconsistent with the recent measurements of the scalar spectral index and tensor-to-scalar ratio. Moreover, apart from this inconsistency, a large symmetry breaking scale is required in the NI model, which may cause large gravitational corrections to the potential. Thus, the NI model should be extended. To this end, the HNI model is an inflation model based on NI. In the HNI model, a low-scale inflation at $f \simeq M$ can be realized by exceeding the value of the running of the scalar spectral index. However, as demonstrated in this study, the symmetry breaking scale should be $f \ge 3.83 M$ in such a model when a negative value of the running of the scalar spectral index is required. Although the positive value of the running of the scalar spectral index is still permissible in the observation, its negative value is favored. 

Overall, this study proposed a generalized HNI model that can realize a low-scale inflation at $f \simeq 0.05 M$. We have shown that the new model is consistent with the observed scalar spectral index and tensor-to-scalar ratio as well as the negative value of the running of the scalar spectral index. If the positive value of the running of the scalar spectral index is excluded from future observations, the validity of the new model would increase. Furthermore, for the completeness of this study, we estimated the running of running of the scalar spectral index, the running of the tensor spectral index, and the energy scale of inflation in the new model.

\vspace{3mm}






\end{document}